\documentclass{aastex}
\usepackage{spr-astr-addons}
\usepackage{url}

\RequirePackage{color}

\begin{document}

\title{Eruptions of Two Coupled Filaments Observed by SDO, GONG and STEREO}
\shorttitle{Eruptions of Two Coupled Filaments}
\shortauthors{Xue et al.}

\author{Z. K. Xue\altaffilmark{1,2}}
 \and \author{X. L. Yan\altaffilmark{1}}
 \and \author{Z. Q. Qu\altaffilmark{1}}
 \and \author{C. L. Xu\altaffilmark{2,3}}
 \and \author{L. Zhao\altaffilmark{1}}

\altaffiltext{1}{Yunnan Observatories, Chinese Academy of Sciences, Kunming 650011, China.}
\altaffiltext{2}{Key Laboratory of Solar Activity, National Astronomical Observatories, Chinese Academy of Sciences, Beijing 100012, China.}
\altaffiltext{3}{Yunnan Normal University, Kunming 650092, China.}

\begin{abstract}
On 2012 July 11, two solar filaments were observed in the northeast of the solar disk and their eruptions due to the interaction between them are studied by using the data from the Solar Dynamics Observatory (SDO), Solar TErrestrial RElations Observatory (STEREO) and Global Oscillation Network Group (GONG). The eastern filament (F1) first erupted toward the northeast. During the eruption of F1, some plasma from F1 fell down and was injected to the North-East part of another filament (F2), and some plasma of F1 fell down to the northern region close to F2 and caused the plasma to brighten. Meanwhile, the North-East part of F2 first started to be active and rise, but did not erupt finally. Then the South-West part of F2 erupted successfully. Therefore, the F2's eruption is a partial filament eruption. Two associated CMEs related to the eruptions were observed by STEREO/COR1. We find two possible reasons that lead to the instability and the eruption of F2. One main reason is that the magnetic loops overlying the two filaments were partially opened by the eruptive F1 and resulted in the instability of F2. The other is that the downflows from F1 might break the stability of F2.
\end{abstract}


\section{Introduction}

Solar filaments (or prominences) are cool and dense plasma embedded in the hot and tenuous corona. They are suspended in the corona and sustained by the magnetic fields which emerge from the photosphere. Filaments often form and exist in filament channels which are along the magnetic polarity inversion lines (PILs). They are classified into three basic types according to their locations: Active region filaments, quiescent filaments and intermediate filaments \citep{eng98}. The trigger mechanism for filament eruption is still unclear, but it is well known that magnetic field plays a key role in filament formation and eruption, such as the magnetic cancellation \citep{sch00,sch06,biy11}, magnetic emergence \citep{che00,yan11}. MHD simulations show that two coronal flux ropes can reconnect when they approach each other \citep{oza97,mil99,kon99,mok01,tor11}. The observation of the filament-carrying flux ropes reconnection was reported by \citet{kum10} and \citet{cha11}. Using H$\alpha$ observations, they found that one filament approached toward another filament, and interacted with each other. Finally, reconnection of the filament-carrying magnetic fields took place. According to \citet{biy12}, the reconnection between the turbulent filament threads and the surrounding magnetic field could occur. \citet{liu12a} reported a partial eruption of a double-decker filament which was composed of two branches. They found that filament threads within the lower branch merged with the upper branch. A lot of theoretical models have been put forward in the past decades to explain the trigger mechanism of filament eruptions, such as the magnetic breakout model \citep{ant99,lyn04,dev08}, the tether-cutting model \citep{moo80,moo01}, the catastrophe model \citep{for90,for91}, and the kink instability model \citep{hoo79,tit99,kli06}. The eruptions of filaments are often accompanied by flares and coronal mass ejections (CMEs) \citep{sub01,yan12a}.

The interactions between filaments and their resulting eruptions have been studied with the observations and simulations in the past decades. \citet{suj07} studied the interaction of two close H$\alpha$ filaments, and their successive eruptions. They concluded that one filament eruption might be triggered by the sudden mass injection from the other one. Three successive, interdependent filament eruptions that took place one by one within 5 hours from different locations were analyzed by \citet{jia11}. They concluded that coronal dimmings might link consecutive eruptions nearby with sympathetic eruptions. \citet{liy12} reported the interactions and eruptions of two filaments using three different angles of view and found that the two filaments were linked together and finally erupted. \citet{yjy12} observed two successive filament eruptions. They suggested that sympathetic eruptions were likely produced by multiple-arcade bipolar helmet-streamer configurations. \citet{she12} investigated a partial and a full eruptions of two solar filaments, and they proposed a possible mechanism within the framework of the magnetic breakout model to interpret the sympathetic filament eruptions. \citet{kon13} found the interaction of two filaments. One leg of first filament swept second filament. The first filament eruption opened the large-scale overlying coronal loops and led to the second filament eruption. Therefore, in one eruptive event including two or multiple filaments, they may interact with each other and finally erupt simultaneously or consecutively.

In this paper, we investigate two filaments and the successive eruptions with multi-wavelength observations from SDO, GONG and STEREO on July 11, 2012, with particular emphasis on the causality chain between the two eruptions. Instruments and data are introduced in Section 2, and the results are presented in Section 3. In Section 4, discussion and conclusion are given.

\section{Observations}

On July 11, 2012, two filaments were observed by space-based (SDO, STEREO) and ground-based telescopes (GONG). We use the multi-wavelength data to study the interaction between the filaments and the resulting successive eruptions. All of the full-disk images taken at different times are rotated to a reference time (08:40 UT) before filament eruptions in order to correct the solar differential rotation. The more detailed information about the data is listed as follows.

The Atmospheric Imaging Assembly \citep[AIA;][]{lem12} on board SDO \citep{pes12} provides multi-channel solar images with high spatial resolution (0.6$\arcsec$ per pixel) and high temporal resolution (12 or 24 s). We use the images in two channels (304 \AA\ and 193 \AA) to show the evolution of the filament eruptions. The Helioseismic and Magnetic Imager \citep[HMI;][]{sch12} is another instrument on board SDO. HMI provides full-disk, high-cadence Doppler, intensity, and magnetograms with a spatial resolution of 1$\arcsec$ (4096$\times$4096-pixel images) of the solar photosphere. At the same time, the full-disk H$\alpha$ line-center images from the ground-based GONG network telescope \citep{har96} with pixel of approximate $1\arcsec$ and 1 minute cadence are used to identify two filaments.

To study the filament eruptions from another angle of view, we also use the 304 \AA\ images obtained by Extreme Ultraviolet Imager \citep[EUVI;][]{how08} telescope on board STEREO-B with a spatial resolution of 1.6$\arcsec$ and a cadence of 10 minutes. The associated CMEs were detected by the COR1 coronagraph \citep{tho03} on board STEREO-B. COR1 is an internally occulted coronagraph. It takes observations of CMEs from 1.3 to 4 solar radii in three different polarizing angles every five minutes.

\section{Results}

\subsection{General characteristics of the two filaments}

On July 11, 2012, two filaments were observed in the northeast of the solar disk. Fig. 1(a) shows the configuration of the two filaments at 08:40:07 UT in AIA 304 \AA\ image before their eruptions. They formed an overall J-shaped structure. The left filament (called F1, hereafter) was closer to solar limb and is marked by a blue arrow in Fig. 1. It had an arch-shaped structure, and its main body was broader than the other one. Observed from H$\alpha$ image at 08:40:14 UT (panel (b)), there were several barbs on the right side of F1. The right filament (called F2) is marked by yellow arrows in panel (a). It appeared to be slender line-shaped and almost perpendicular to F1 projectively. Panel (c) shows the magnetic configuration where the filaments were located. The magnetic configuration is composed of many discrete flux elements which can be classified into two groups. One group is dominated by positive polarities and the other group by negative polarities. Negative magnetic fields are mainly distributed around the positive ones. The contours of the positive and negative magnetic fields are overlaid on the GONG/H$\alpha$ image (see panel (b)) with white and black curves respectively. One can see that the filaments were nearly seated along the magnetic polarity inversion line. Before the filament eruptions, no flare was detected at the same position.

\subsection{The filament eruptions}

Fig. 2 shows the snapshots of F1 evolution observed in AIA 304 \AA\ (panels (a)-(f)), GONG/H$\alpha$ (panels (g)-(i)) where F1 is indicated by the white arrows. Before its eruption, F1 showed a stable structure in its initial position in both AIA 304 \AA\ and GONG/H$\alpha$ images (panels (a) and (g)). At about 08:50 UT, F1 began to rise toward the solar northeast slowly and departed from F2 gradually. The apex of F1 expanded more quickly than the flank seen from the AIA 304 \AA\ images. From about 09:20 UT, F1 ascended more quickly. To distinguish F1 from the background, we present a series of AIA 304 \AA\ base difference images (panels (d)-(f)) to display the evolution, which are produced by subtracting the image observed at 08:40:07 UT from each image. Note that the white patch represents  the positions of filaments before their eruptions and the black patch represents the erupting filaments in base difference images. It was found that F1 erupted to the northeast until 10:52 UT and it disappeared completely in AIA 304 \AA\ base difference images after 10:52 UT (panel (f)). In the H$\alpha$ images, F1's behavior was similar to that observed in AIA 304 \AA\ channel, but it disappeared quickly at about 09:45 UT (panel (i)).

During F1 eruption, one part of the F1 plasma fell down  along the right leg of F1, formed downflows and caused brightening. A similar scenario was observed by \citet{inn12} and \citet{xue14}. Fig. 3 shows the evolutions of the downflows (indicated by green arrows) and brightening (indicated by blue arrows) in AIA 304 \AA\ images (panels (a)-(i)) and 171 \AA\ images (panels (j)-(l)). At the beginning, the downflows exhibited bright, threadlike shapes. It seems that the downflowing plasma was injected into the North-West part of F2 (panels (a)-(b)). With the rising of F1, the downflows moved to the northwest gradually and the end points of downflows moved from F2 to the northern region (panels (c)-(f)). When the downflows reached the lower atmosphere, it caused the plasma to brighten. The brightening first appeared in the region close to F2 and became brighter and brighter gradually with the mass of falling plasma, which can also be seen clearly in AIA 171 \AA\ channel.

To study the relationship between the downflows, brightening, and the eruption of F2 in detail, along the white lines marked by S1 and S2 in Fig. 4(a), we obtain two time-slice maps using AIA 304 \AA\ images and display them in Fig. 4(b) and (c) respectively. S1 is along the path of the downflows, and S2 along the main body of F2. In panel (b), the downflows show several bright diagonal structures pointed by green arrow. During the eruption of F1 (indicated by white arrow and white dashed line), the downflowing plasma fell down to F2 (indicated by black arrow and yellow dashed line) and the lower atmosphere. We also calculate the average speed of the downflows projected onto the plane of the sky by fitting the timeslice map along S1 with linear functions, and the fitting result are showed by dashed lines. The average speed is about 68.8 km s$^{-1}$, close to the results obtained by other authors \citep{liu12b,tri06}. Panel (b) also shows that the brightening appeared when downflows reached the lower solar atmosphere. The brightening was very close to F2. Panel (c) displays the timeslice map along S2. During the downflows and brightening, the North-West part of F2 started to activate at around 09:00 UT. We can see that the plasma in North-West part of F2 moved from right to left until the South-West part of F2 erupted. In timeslice map (panel (c)), the moving plasma exhibits displacements (indicated by white dashed lines) from the lower left to the upper right. The intensity over the area marked by the blue rectangle in Fig. 4(a) is integrated to investigate the evolution of the intensity of brightening. The intensity-time profile normalized by the intensity at 08:40 UT is obtained from AIA 304 \AA\ images and shown in Fig. 4(d). Because of the downflows, the intensity started to increase quickly from about 09:07 UT to 09:31 UT, then the increase became slow until its first peak value (P1) at 09:54 UT. After P1, it decreased to its minimum value at 10:16 UT. We also find that there is another peak value (P2) at about 10:40 UT, caused by the activity of North-West part of F2, and was not related to the downflows and the main ribbons led by filament eruptions. During the first increasing stage (from 09:07 UT to 10:16 UT), the North-West part and South-West part of F2 started to rise at 09:18 UT (dotted line) and 09:50 UT (dashed line) respectively.

Fig. 5 displays the process of F2 eruption. Panels (a)-(c) show the AIA 304 \AA\ images and panels (d)-(f) are base difference images in the AIA 304 \AA\ channel. The GONG/H$\alpha$ images are shown in panels (g)-(i) and the AIA 193 \AA\ images in panels (j)-(l). At about 09:10 UT, the North-West part of F2 first started to be active,  and rose slowly from 09:18 UT. Then the AIA 304 \AA\ images show that, at about 09:50 UT, the South-West part of F2 started to erupt toward the northeast. At the same time, the North-West part of F2 rose to a certain height but did not erupt finally. It indicates that F2 erupted partially. The South-West part of F2 was visible (marked by the white arrow in panels (d)-(e)) until it went off the solar disk (panel (f)). The eruption of F2 was also visible in H$\alpha$ images. The South-West part of F2 rose and disappeared gradually(panels (h) and (i)), but the North-West part of F2 can be still observed. During this process, two bright ribbons were observed on both sides of erupting part of F2 and marked by two black arrows (panels (h)-(i)). After the eruption of F2, we observe a series of bright post flare loops at the position of F2 in AIA 193 \AA\ images (panel (l)). The ribbons in H$\alpha$ images were located at the footpoints of the post flare loops (indicated by black arrows).

During the eruption, two dimming regions appeared and were most clearly observed in AIA 193 \AA\ images. Figure 4(k) shows the positions of the two dimming regions with two black rectangles, marked by symbols ``D1" and ``D2" respectively, and the position of F2 before eruption is indicated by white line. The two dimming regions first appeared at the onset of F2 eruption. D1 was located at the middle of F2, exactly corresponding to the left of erupting part of F2. D2 was located at the right footpoint of F2. This implies that the two dimming regions were caused by the F2 eruption. Generally, it is believed that the dimmings represent the footprints of a large-scale flux-rope ejection \citep{jia07,jia11}. We conclude that the dimming regions were caused by the expansion of the F2 flux rope and the subsequent mass depletion.

\subsection{Filament eruptions and CMEs observed by STEREO-B}

To show this event from another perspective, we display wide field-of-view (FOV) images observed by STEREO-B/EUVI 304 \AA\ channel and COR1 respectively. Fig. 6 presents the eruptive filaments (panels (a1)-(a3)) and associated CMEs (panels (b1)-(b3)). The eruptive filaments can be seen clearly in STEREO-B/EUVI 304 \AA\ images and are indicated by the white arrows. In panel (a1), F1 erupted to the northwest with a rectangle shape. At 10:26 UT, F1 erupted away from the solar limb, and its main body can be seen more clearly. At the same time, F2 can be also seen off the solar limb. In panel (a3), F1 erupted and then disappeared from EUVI 304 \AA\ images. F2 showed an arched shape and finally erupted.

Meanwhile, two CMEs were observed by STEREO-B/COR1. Fig. 6(b1)-(b3) display the CMEs with base difference images of STEREO/COR1. The base difference images are produced by subtracting the image observed at 08:50:34 UT by the following images. The CMEs are marked by ``CME1" and ``CME2" respectively. CME1 first appeared in the COR1 images at 09:35 UT and propagated to the northwest with a fan-shaped structure. COR1 detected CME2 firstly at 10:30 UT and its shape was similar to that of CME1, but it was more slender and its propagation path was closer to the solar equator. According to the onsets and the positions of the CMEs, we conclude that CME1 was connected to F1 eruption and CME2 to F2 eruption. They propagated along different paths which were consistent with the paths of filament eruptions.

\subsection{Coronal magnetic configuration}

To study this erupting event in detail, we investigate the topology of coronal magnetic fields using the potential field source surface \citep[PFSS;][]{sch03} model. Fig. 7 presents different views of the same representative PFSS coronal magnetic field lines which originate from the regions around the two filaments. F1 and F2 are outlined with blue and yellow color respectively. The PFSS extrapolation is indicated by the red curves. Fig. 7(a) shows the locations of the two filaments as the same time and same FOV as Fig. 1(a). One can see the magnetic field lines laying over the magnetic polarity inversion line, and linking one magnetic polarity to the opposite magnetic polarity. The two footpoints of each magnetic loop are located on different side of the filaments. Fig. 7(b) has the same time and same FOV as Fig. 6(a1). The two filaments are located at the solar limb, therefore we can compare the height of filaments and that of magnetic loops. It shows that the magnetic loops are higher than the filaments. This implies that the filaments were located below the magnetic field lines. Meanwhile, we also find that a few of magnetic field lines overlying F1 also were shared by F2.

\section{Conclusion and Discussion}

We present the observations of the eruptions of two filaments which were observed by three different instruments (STEREO-B, SDO, and GONG) in multi-channels, such as 304 \AA, 193 \AA, H$\alpha$ and white light. This event contains several interesting aspects. The major observational facts in the observational analysis can be summarized as follows. Two filaments were mainly located above the magnetic polarity inversion line. One filament F1 first erupted to the northeast. During F1 eruption, a part of F1 plasma fell down and was injected into F2, besides that causing plasma brightenings. Then the North-West part of F2 started to be active, but it did not erupt. The South-West part of F2 started to rise and erupted successfully. During the F2 eruption, two dimming regions were observed. After the filament eruptions, STEREO-B/COR1 detected two associated CMEs propagating in the corona along two different paths.

It is suggested by many authors that the magnetic arcades overlying filaments could be removed by other solar eruptive events \citep{web97,gar04,nag07,jia09,yan12b,kon13}. \citet{zuc09} proposed a so-called domino effect:  a first filament eruption caused a lift-off of the inner arcade and resulted in the destabilization and eruptions of other two filaments. The eruptions of three sympathetic filaments connected by coronal dimmings were studied by \citet{jia11}. They found that a filament caused a weakening and partial removal of an overlying magnetic field of the two other filaments that triggered them to erupt. The paper of \citet{yjy12} showed that the simple expansion of the common streamer arcade forced by the first filament eruption weakened magnetic loops overlying another filament and thus led to its eruption.

The configuration of magnetic field lines plays an important role in the stabilization of filaments. Once the stability of the magnetic field is destroyed, the filaments may be active or erupt. Based on the magnetic configuration shown in Fig. 7, we propose that the magnetic loops overlying F1 were pushed outward and finally removed or open due to rise of F1 in the process of F1's eruption. Since some magnetic loops overlying F1 were also located above another filament F2, the erupting F1 pushed outward those magnetic loops. As a result, the overall magnetic field configuration of F2 was destroyed and this causes the instability of F2. Our suggestion is supported by the fact that the North-West part of F2 first started to be unstable and dynamic. Finally, F2 erupted partially.

When more than one filaments are located close to each other, if one became unstable, it may trigger others to be active. They may interact with each other and finally erupt. \citet{suj07} observed that a filament was triggered to erupt by the nearby erupting filament. They concluded that the second filament eruption was caused by an overload of mass injecting suddenly from the first filament. \citet{liu10} found two filaments to interact with each other, but the filaments erupted unsuccessfully. They reported that the two filaments merged together along the ejection path and deduced that the bodily coalescence between the two interacting flux ropes took place. The event that two filaments were linked together, interacted with each other, and finally erupted was observed by \citet{liy12} from different viewing angles. \citet{bon09} reported an interesting event occurred on 19 May 2007. Before the eruptions, an active region filament and a quiescent filament formed, interacted with each other and merged. Finally, the merged filament erupted and caused a complex CME.

In our event, two filaments erupted successively along two different paths. F1 first erupted at 08:50 UT. During F1 eruption, we observe some plasma from F1 was injected into the North-West part of F2, and at the same time (09:00 UT), the plasma in F2 moved from right to left along the F2. Meanwhile, another part of plasma in F2 fell back to the northern region of F2 and caused brightening. The brightening started at 09:07 UT and was very close to F2. Then the North-West part of F2 became active at around 09:18 UT. When the brightening became brighter, close to the maximum value at 09:50 UT, the South-West part of F2 started to rise and erupted finally. Based on the temporal and spatial relationships of the downflows, brightening and F2, therefore, we conclude that the downflows from F1 might disturb F2 and caused F2 to erupt finally. Evidently, this event is similar as the observations of Su et al. (2007) where the mass from the first erupting filament was injected to the second filament and caused the second filament to erupt.

In summary, based on the multi-channel observations of SDO, GONG and STEREO and the magnetic field topology, we study the relationship and interaction between two coupled filaments and their eruptions. Based on the above discussions, we suggest that two filament eruptions were associated with each other rather than independent and conclude two possible reasons for the F2's eruption. One reason is that a part of the magnetic loops overlying F2 were removed or opened by the F1's eruption because the magnetic loops were shared by the two filaments. Another one is that the second filament eruption may be due to the downflows from F1 which may injected into F2 partly and caused the loss of stability of F2. As a result, F2 was triggered to activate and erupted finally.

\begin{figure*}
\centering
   \includegraphics[width=\textwidth]{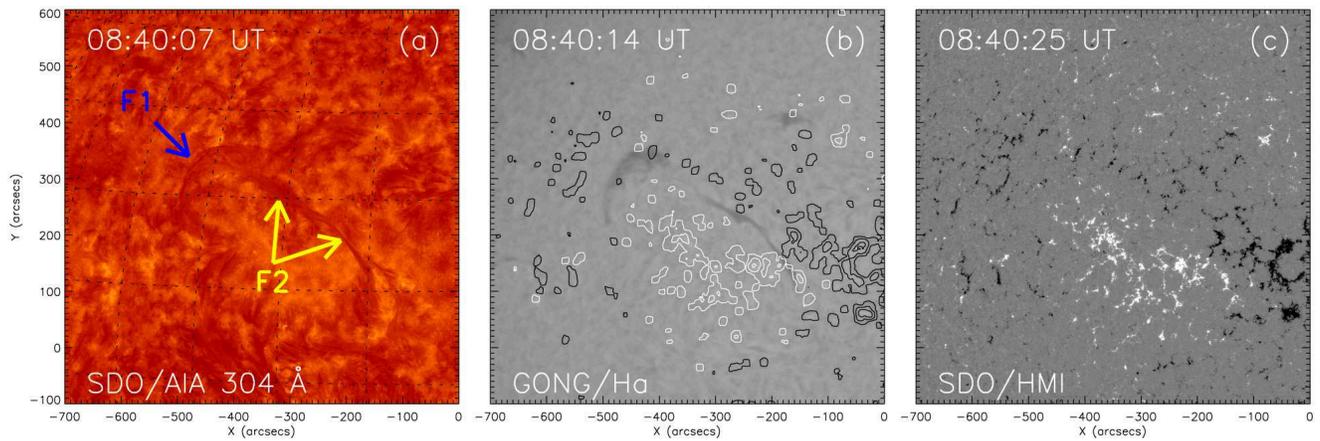}
\caption{The positions of two filaments before their eruptions. Panel (a) shows two filaments (indicated by different color arrows) in AIA 304 \AA\ image. Panel (b) displays GONG/H$\alpha$ image, overlaid with contours ($\pm$20, $\pm$100, $\pm$250 G) from the line-of-sight magnetogram shown in panel (c) obtained by HMI. Negative/positive polarity contours are in black/white color curves. }
\end{figure*}

\begin{figure*}
\centering
   \includegraphics[width=\textwidth]{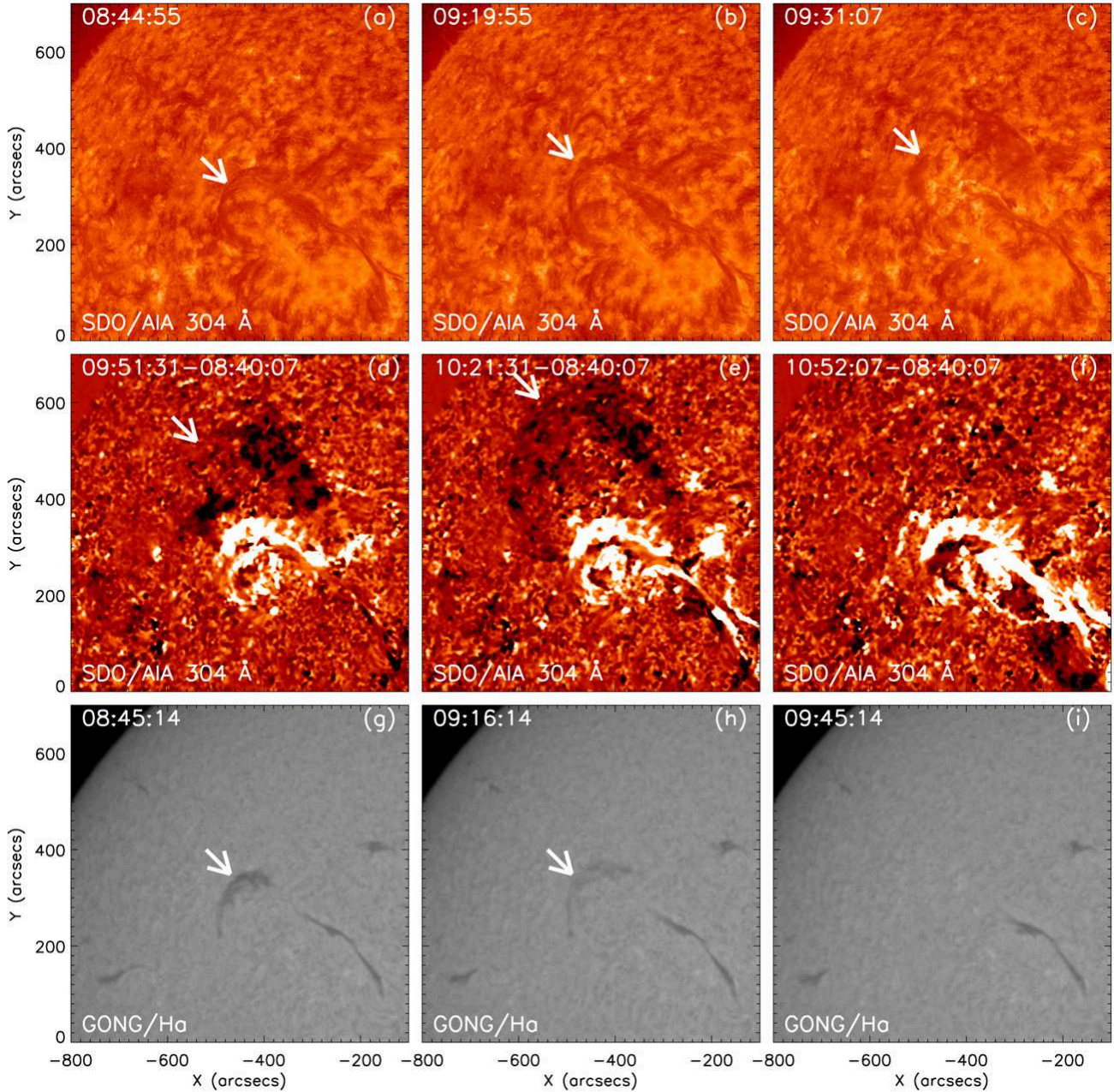}
\caption{The process of the first filament eruption. Panels (a) - (c) show AIA 304 \AA\ images. Panels (d)-(f) display AIA 304 \AA\ base difference images. GONG H$\alpha$ images are shown in panels (g)-(i). The position of the first filament is marked by the white arrows.  }
\end{figure*}

\begin{figure*}
\centering
   \includegraphics[width=\textwidth]{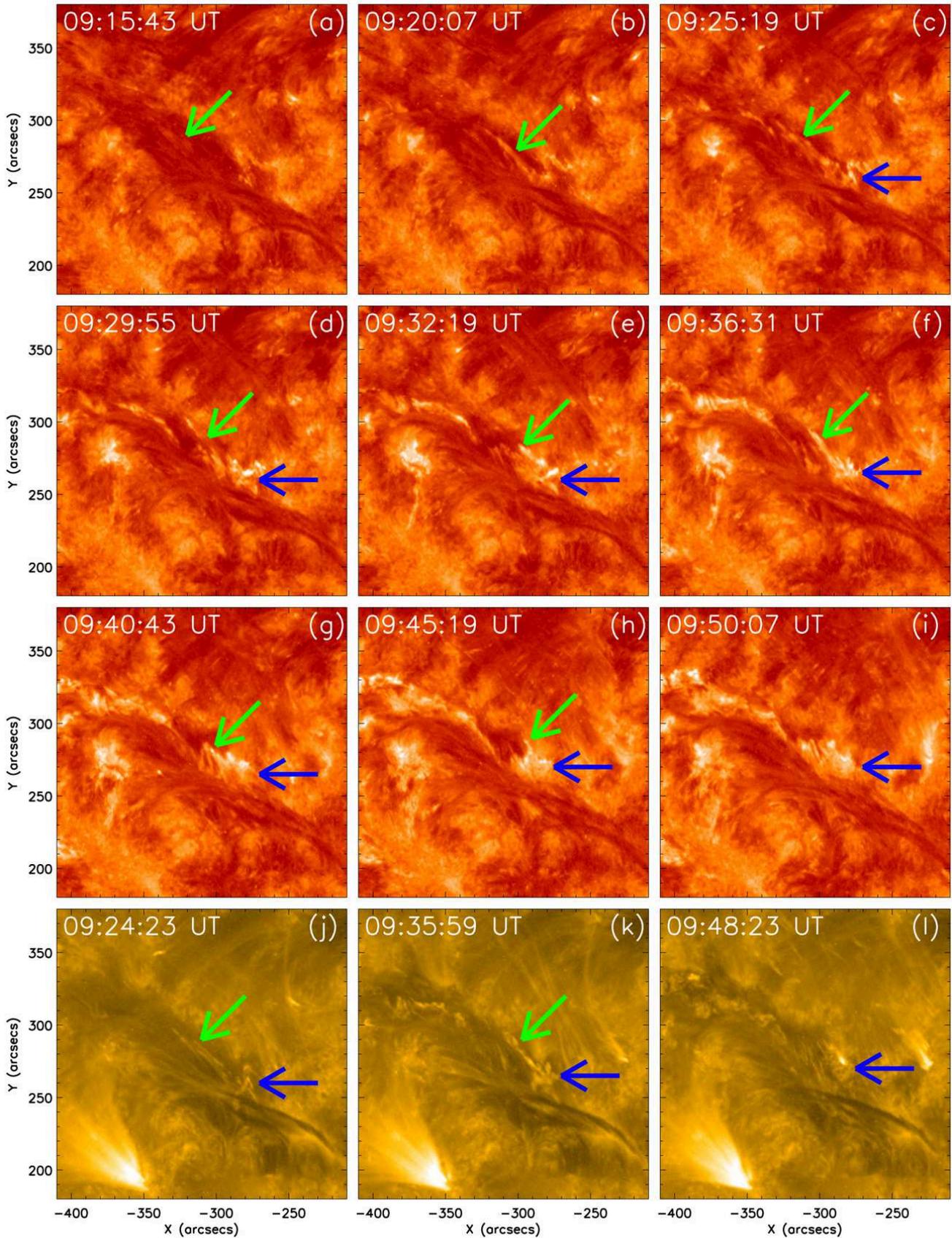}
\caption{The evolutions of the downflows and brightening marked by the green and blue arrows respectively are shown in AIA 304 \AA\ images (panels (a)-(i)) and 171 \AA\ images (panels (j)-(l)). }
\end{figure*}

\begin{figure*}
\centering
   \includegraphics[width=\textwidth]{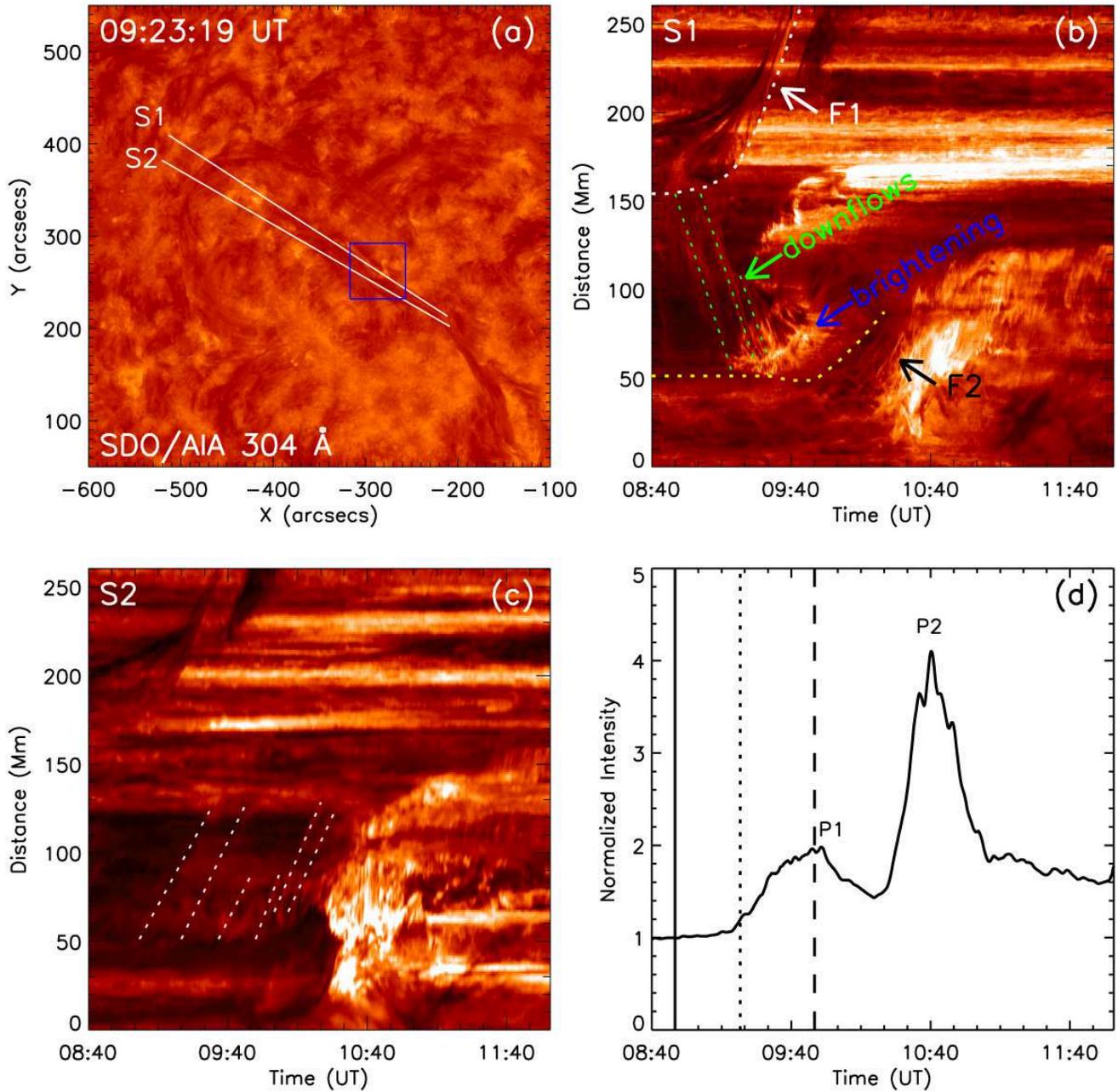}
\caption{The positions of two slices (S1, S2) are shown in AIA 304 \AA\ image (panel (a)) and time-slice maps (shown in panels (b) and (c)) are calculated along the two slices respectively. In panel (b), F1 and F2 are indicated by arrows and white and yellow dashed lines. The green and blue arrows point to the downflows and brightening. The paths of downflows are marked by green dashed lines. In panel (c), the moving features in F2 are marked by white dashed lines.The intensity-time profile of AIA 304 \AA\ images is calculated in the area indicated by the blue rectangle (panel (a)) and is shown in panel (d). The profile is normalized to the intensity at 08:40 UT. The solid line indicates the starting time of F1's eruption. The dotted line and dashed line denote the onsets of eruptions of the North-West part and South-West part of F2 respectively. }
\end{figure*}

\begin{figure*}
\centering
   \includegraphics[width=\textwidth]{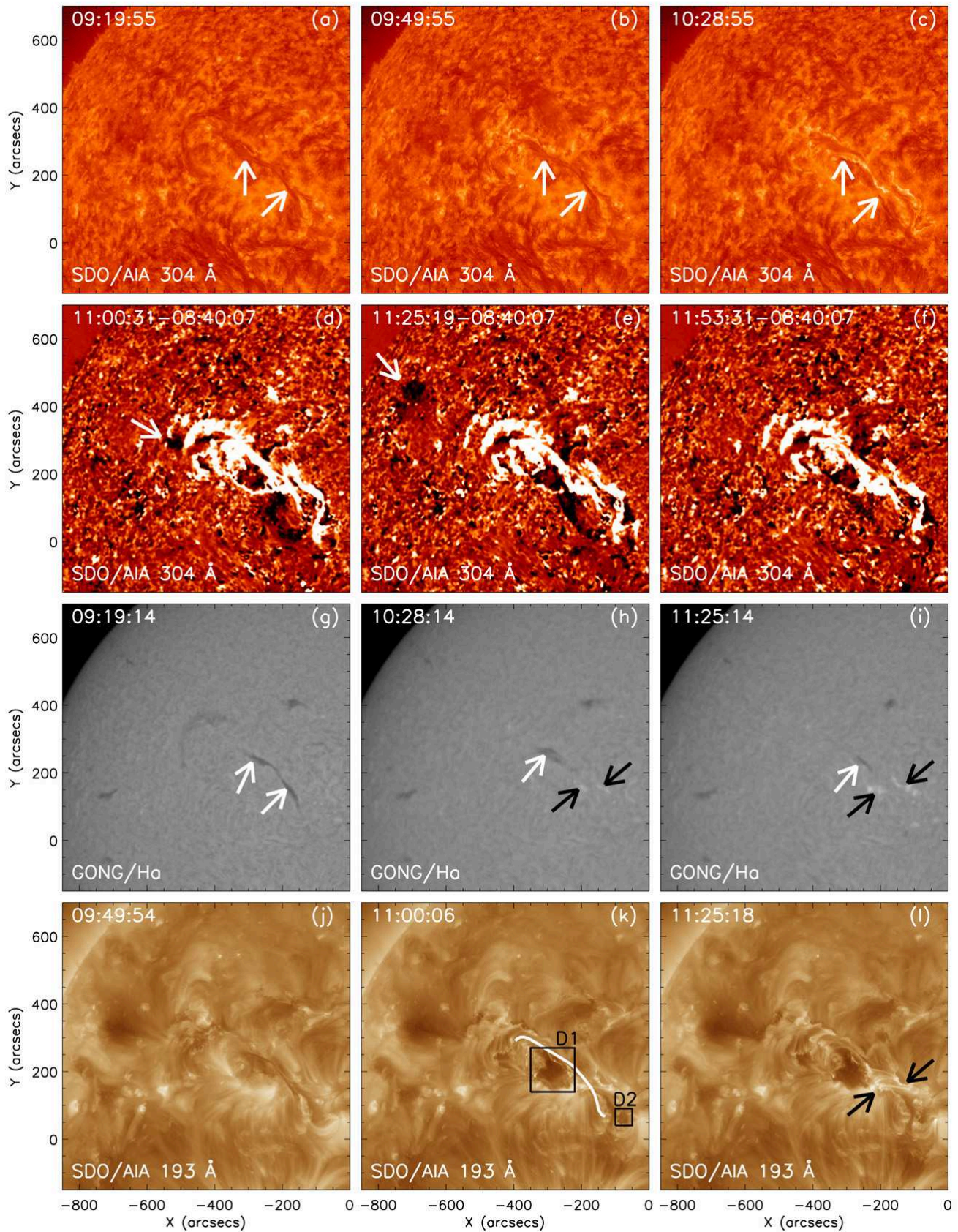}
\caption{The different wavelength images to show the eruption of F2. Panels (a)-(c) are the AIA 304 \AA\ images. Panels (d)-(f) are the base difference images of AIA 304 \AA. Panels (g)-(i) are the GONG/H$\alpha$ images and panels (j)-(l) are the AIA 193 \AA\ images. The position of F2 is marked by the white arrows. The two bright ribbons are indicated by the black arrows in panels (h), (i) and (l). The white line in panel (k) represents the position before its eruption.}
\end{figure*}

\begin{figure*}
\centering
   \includegraphics[width=\textwidth]{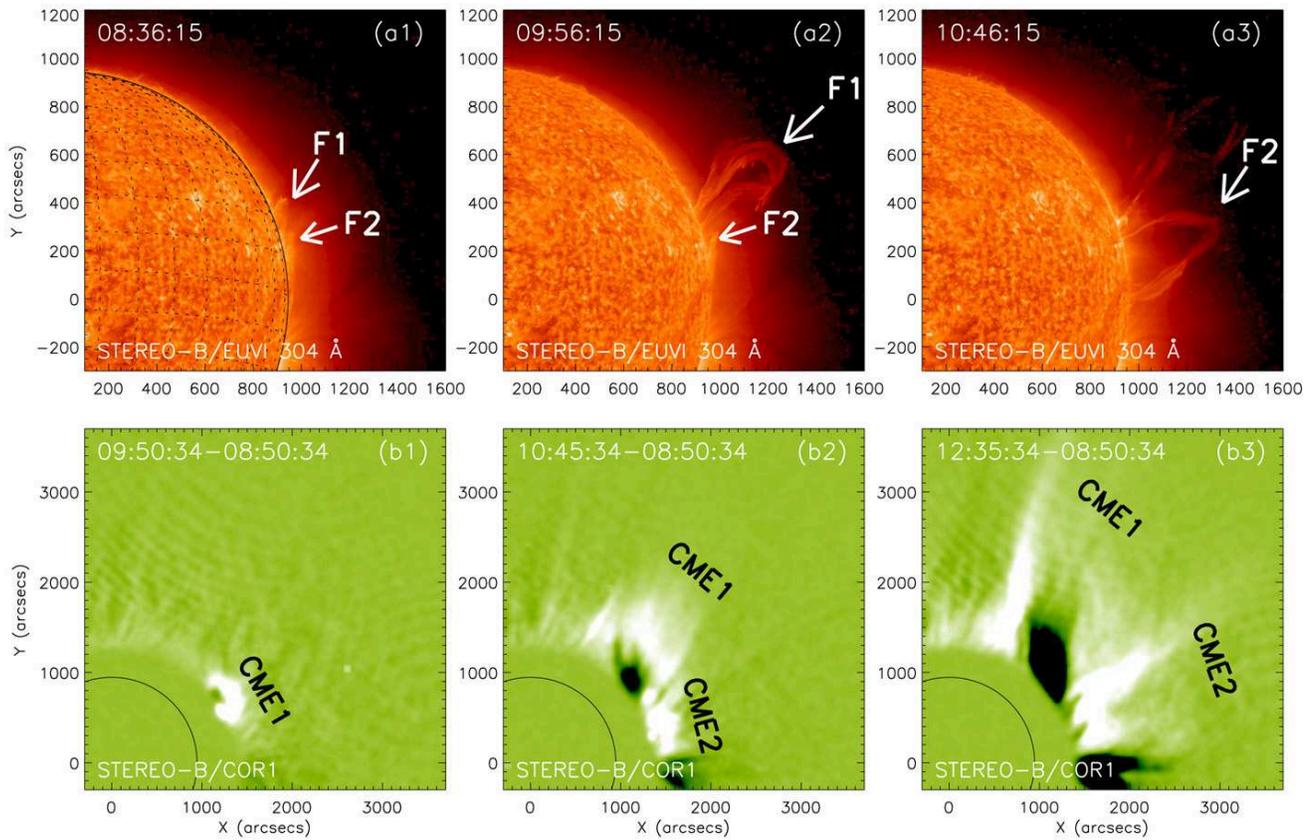}
\caption{Filament eruptions (panels (a1)-(a3)) and CMEs (panels (b1)-(b3)) observed with STEREO-B/EUVI 304 \AA\ channel and COR1. The filaments are indicated by white arrows and two CMEs are marked by `CME1' and `CME2' respectively. }
\end{figure*}

\begin{figure*}
\centering
   \includegraphics[width=\textwidth]{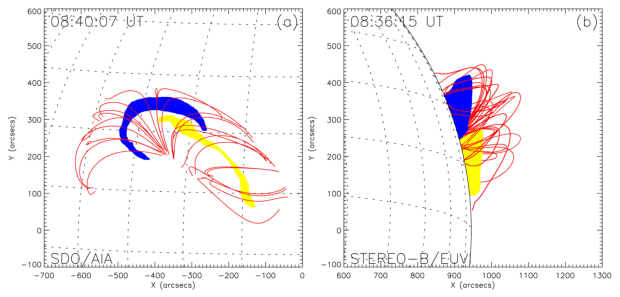}
\caption{The magnetic field lines (indicated by red curves) extrapolated from PFSS model viewed differently from SDO (panel (a)) and STEREO-B (panel (b)). F1 and F2 are outlined with blue and yellow color respectively. }
\end{figure*}

\acknowledgments
We would like to thank the referee for many valuable and insightful comments which greatly helped us to improve the manuscript. We also thank the SDO, GONG and STEREO teams for data support. This work is sponsored by the National Science Foundation of China (NSFC) under the grant numbers 11373066, 11373065, 11103075, Key Laboratory of Solar Activity of CAS under number KLSA201412, KLSA201212, Yunnan Science Foundation of China under number 2013FB086, 2013FZ041, and the Talent Project of Western Light of Chinese Academy of Sciences.


\begin{thebibliography}{}

\bibitem[Antiochos et al.(1999)]{ant99} Antiochos, S.K., DeVore, C.R., Klimchuk, J.A., 1999. ApJ 510, 485.

\bibitem[Bi et al.(2011)]{biy11} Bi, Y., Jiang, Y. C., Yang, L. H., et al., 2011, \na, 16, 276

\bibitem[Bi et al.(2012)]{biy12} Bi, Y., Jiang, Y. C. Li, H. D., et al., 2012, \apj, 758, 42

\bibitem[Bone et al.(2009)]{bon09} Bone, L. A., van Driel-Gesztelyi, L., Culhane, J. L., et al., 2009, \solphys, 259, 31

\bibitem[Chandra et al.(2011)]{cha11} Chandra, R., Schmieder, B., Mandrini, C. H., et al., 2011, \solphys, 269, 83

\bibitem[Chen \& Shibata(2000)]{che00} Chen, P. F., \& Shibata, K., 2000, \apj, 545, 524

\bibitem[DeVore \& Antiochos(2008)]{dev08} DeVore, C.R., \& Antiochos, S.K., 2008, \apj, 680,740.

\bibitem[Engvold(1998)]{eng98} Engvold, O., 1998, in ASP Conf. Ser. 150, IAU Colloq. 167: New Perspectives on Solar Prominences, ed. D. F. Webb, B. Schmieder, \& D. M. Rust (San Francisco, CA: ASP), 23

\bibitem[Forbes(1990)]{for90} Forbes, T. G., 1990, \jgr, 95, 11919

\bibitem[Forbes \& Isenberg(1991)]{for91} Forbes, T.G., \& Isenberg, P.A., 1991, \apj, 373, 294.

\bibitem[Gary \& Moore(2004)]{gar04} Gary, G. A., \& Moore, R. L., 2004, \apj, 611, 545

\bibitem[Harvey et al.(1996)]{har96} Harvey, J. W., Hill, F., Hubbard, R. P., et al., 1996, Science, 272, 1284

\bibitem[Hood \& Priest(1979)]{hoo79} Hood, A.W., \& Priest, E.R., 1979, \solphys, 64, 303.

\bibitem[Howard(2008)]{how08} Howard, R. A., Moses, J. D., Vourlidas, A., et al., 2008. \ssr, 136, 67

\bibitem[Innes et al.(2012)]{inn12} Innes, D.E., Cameron, R.H., Fletcher, L., Inhester, B., Solanki, S.K., 2012, \aap 540, L10

\bibitem[Jiang et al.(2007)]{jia07}  Jiang, Y. C., Yang, L. H., Li, K. J., Ren, D. B., 2007, \apj, 662, L131

\bibitem[Jiang et al.(2009)]{jia09} Jiang, Y., Yang, J., Zheng, R., et al., 2009, \apj, 693, 1851

\bibitem[Jiang et al.(2011)]{jia11} Jiang Y. C., Yang J. Y., Hong, J. C., et al., 2011, \apj, 738, 179

\bibitem[Kliem \& T\"{o}r\"{o}k(2006)]{kli06} Kliem, B., \& T\"{o}r\"{o}k, T., 2006, \prl, 96, 255002.

\bibitem[Kondrashov et al.(1999)]{kon99} Kondrashov, D., Feynman, J., Liewer, P. C., Ruzmaikin, A., 1999, \apj, 519, 884

\bibitem[Kong et al.(2013)]{kon13} Kong, D. F., Yan, X. L., \& Xue, Z. K., 2013, \apss, 348, 303

\bibitem[Kumar et al.(2010)]{kum10} Kumar, P., Manoharan, P. K. \& Uddin, W., 2010, \apj, 710, 1195

\bibitem[Lemen et al.(2012)]{lem12} Lemen, J. R., Title, A. M., Akin, D. J. et al., 2012, \solphys, 275, 17

\bibitem[Li \& Ding(2012)]{liy12} Li Y., \& Ding M. D., 2012, RAA, 12, 3

\bibitem[Liu et al.(2010)]{liu10} Liu, Y., Su, J., Shen, Y., \& Yang, L., 2010, in IAU Symposium 264, eds. A. G. Kosovichev, A. H. Andrei, \& J.-P. Roelot, 99

\bibitem[Liu et al.(2012a)]{liu12a} Liu, R., Kliem, B., T\"{o}r\"{o}k, T., et al., 2012a, \apj, 756, 59

\bibitem[Liu et al.(2012b)]{liu12b} Liu W, Berger, T., Low, B.C., 2012b, \apj, 745, L21

\bibitem[Lynch et al.(2004)]{lyn04} Lynch, B. J., Antiochos, S. K., MacNeice, P. J.,et al., 2004, \apj, 617, 589

\bibitem[Milano et al.(1999)]{mil99} Milano, L. J., Dmitruk, P., Mandrini, C. H., et al., 1999, \apj, 521, 886

\bibitem[Mok et al.(2001)]{mok01} Mok, Y., Miki\'{c}, Z., Linker, J., 2001, \apj, 555, 440

\bibitem[Moore \& LaBonte(1980)]{moo80} Moore, R.L., \& LaBonte, B.J., 1980, in IAU Symp. 91, Solar and Interplanetary Dynamics, ed. M. Dryer \& E. Tandberg-Hanssen (Dordrecht: Reidel), 207.

\bibitem[Moore et al.(2001)]{moo01} Moore, R.L., Sterling, A.C., Hudson, H.S., Lemen,J.R., 2001, \apj, 552, 833.

\bibitem[Nagashima et al.(2007)]{nag07} Nagashima, K., Isobe, H., Yokoyama, T., et al., 2007, \apj, 668, 533

\bibitem[Ozaki \& Sato(1997)]{oza97} Ozaki, M., \& Sato, T., 1997, \apj, 481, 524

\bibitem[Pesnell et al.(2012)]{pes12} Pesnell, W. D., Thompson, B. J., Chamberlin, P. C., 2012, \solphys, 275, 3

\bibitem[Scherrer et al.(2012)]{sch12} Scherrer, P. H., Schou, J., Bush, R. I., et al., 2012. \solphys, 274, 229

\bibitem[Schmieder et al.(2000)]{sch00} Schmieder, B., Delann\'{e}e, C., Deng, Y. Y., et al., 2000, \aap, 358, 728

\bibitem[Schmieder et al.(2006)]{sch06} Schmieder, B., Aulanier, G., L\'{o}pez Ariste, A., et al., 2006, \solphys, 238, 245

\bibitem[Schrijver \& De Rose(2003)]{sch03} Schrijver, C. J. \& De Rose, M. L., 2003, \solphys, 212, 165

\bibitem[Shen et al.(2012)]{she12} Shen Y. D., Liu Y., Su J. T., 2012, \apj, 750, 12

\bibitem[Su et al.(2007)]{suj07} Su J. T., Liu, Y., Kurokawa, H. et al., 2007, \solphys, 242, 53

\bibitem[Subramanian \& Dere(2001)]{sub01} Subramanian, P., \& Dere, K. D., 2001, \apj, 561, 372

\bibitem[Thompson et al.(2003)]{tho03} Thompson, W. T., Davila, J. M., Fisher, R. R., et al., 2003, Proc. SPIE, 4853, 1

\bibitem[Titov \& D\'{e}moulin(1999)]{tit99} Titov, V.S., \& D\'{e}moulin, P., 1999, \aap, 413, 27.

\bibitem[T\"{o}r\"{o}k et al.(2011)]{tor11} T\"{o}r\"{o}k, T., Chandra, R. Pariat, E., et al., 2011, \apj, 728, 65

\bibitem[Tripathi et al.(2006)]{tri06} Tripathi, D., Isobe, H., Mason, H.E., 2006, \aap, 453, 1111

\bibitem[Webb et al.(1997)]{web97} Webb, D. F., Kahler, S. W., McIntosh, P. S., \& Klimchuck, J. A., 1997, \jgr, 102, 24161

\bibitem[Xue et al.(2014)]{xue14} Xue, Z. K., Yan, X. L., Qu, Z. Q., et al., 2014, \na, 26, 23.

\bibitem[Yan et al.(2011)]{yan11} Yan, X. L., Qu, Z. Q., Kong, D. F., 2011. \mnras, 414, 2803

\bibitem[Yan et al.(2012a)]{yan12a} Yan, X. L., Qu, Z. Q., Kong, D. F., et al., 2012a. \apj, 754, 16

\bibitem[Yan et al.(2012b)]{yan12b} Yan, X. L., Xu, C. L., Qu, Z. Q., \& Kong, D. F., 2012b. \apss, 341, 231

\bibitem[Yang et al.(2012)]{yjy12} Yang J. Y., Jiang, Y. C., Zheng R. S., et al., 2012, \apj, 745, 9

\bibitem[Zuccarello et al.(2009)]{zuc09} Zuccarello, F., Romano, P., Farnik, F., et al., 2009, \aap, 493, 629


\end{thebibliography}
\end{document}